\documentclass[twocolumn,superscriptaddress,aps,pra]{revtex4-1}
\usepackage{ulem,bm}
\usepackage{amsmath, upgreek, amssymb, graphicx, paralist, gensymb,epstopdf}
\usepackage[colorlinks, linkcolor=blue, citecolor=blue, urlcolor=blue]{hyperref}
\usepackage{array,epsfig,amsmath,amssymb}
\usepackage{graphicx} 
\usepackage{dsfont}
\usepackage{color}

\def\bra#1{\langle #1|}
\def\ket#1{|#1 \rangle}

\begin{document}
\title{Noise-adaptive test of quantum correlations with quasiprobability functions}

\author{Seung-Woo Lee}
\email{swleego@gmail.com}
\affiliation{Quantum Universe Center, Korea Institute for Advanced Study, Seoul 02455, Korea}

\author{Jaewan Kim}
\affiliation{School of Computational Sciences, Korea Institute for Advanced Study, Seoul 02455, Korea}

\author{Wonmin Son}
\email{sonwm71@gmail.com}
\affiliation{Department of Physics, Sogang University, Seoul 04107, Korea}

\date{\today}

\begin{abstract}
We introduce a method for testing quantum correlations in terms of quasiprobability functions in the presence of noise. We analyze the effects of measurement imperfection and thermal environment on quantum correlations, and show that their noise effects can be well encapsulated into the change of the order parameter of the generalized quasiprobability function. We then formulate a noise-adaptive entanglement witness in the form of a Bell-type inequality by using the generalized quasiprobability function. Remarkably, it allows us to observe quantum correlations under severe noise. Our method provides a useful tool to test quantum correlations in near-term noisy quantum processors with continuous variable systems.
\end{abstract}

\maketitle

\section{Introduction}

Quantum correlations are essential resources for quantum computation \cite{QIQM}, communication \cite{Bennett93,CVT1,Pirandola2015,Hensen15} and cryptography \cite{Ekert91}. Such quantum features, however, are fragile in the presence of noise in the system or measuring device, and the noise effects tend to become more severe as the size of the system increases. One of the most important challenges in building scalable quantum processor \cite{SQArchi1,SQArchi2} and network \cite{Kimble08, Pan17, Wehner18, SWL19, SML19} is thus an efficient identification of quantum correlations in noisy quantum systems \cite{NISQ}.

Meanwhile, a continuous variable (CV) system can be effectively represented by quasiprobability functions in phase space. The quasiprobability distribution such as the Glauber-Sudarshan $P$-function \cite{Sudarshan,Glauber63}, the Wigner function \cite{Wigner}, and the Husimi $Q$-function \cite{Husimi} can be used as an equivalent description to the density matrix \cite{Cahill69a,Cahill69b,Moya93}. Nonclassical features of quantum states, which are incompatible with classical counterparts, such as the negativity \cite{Dahl06,Walschaer17,Bohmann20}, nonlocality \cite{Banaszek99,SWLee09}, contextuality \cite{Spekkens08,Asadian15}, entanglement \cite{MSK02,JS09,JS19l} and coherence \cite{JS18} were studied in this formalism as useful resources for quantum information processing \cite{Veitch12,Pashayan15,Keshari16,Shahandeh17,JS19a,Weinbub18}. Verifying quantum correlations in CV systems has thus been an important issue in quantum technologies \cite{CVQI1,CVQI2,CVQI3}. 

Methods to directly measure the quasiprobability function of a given quantum state has been actively developed \cite{Banaszek96,Wallentowitz96,Banaszek99a,Banaszek02,Bertet02,Juarez-Amaro03}, while a tomography by homodyne detection has been typically used to reconstruct the quasiprobability distribution \cite{pndist,Leonhardt93,Lvovsky2001,Thomas00,HWLee15}. If the field can be confined in a cavity, the quasiprobability function is directly measurable by the atom-field interaction because it provides useful means to detect cavity quantum electrodynamic systems \cite{Bertet02,Juarez-Amaro03}. Recent progress of the photon-number-resolving detectors \cite{Eisamana11}, e.g.,~based on superconducting circuits \cite{Schuster07,Hofheinz09,Marsili09}, enhances the possibility to directly measure the quasiprobability function as well as the nonclassical features of a quantum state \cite{Luis15,Bohmann18}. However, such a measurement is very sensitive to the noise in the measured system or the measuring device. If the measurement imperfection or environmental noise become significant, observed quantum features tend to disappear sharply. Entanglement witnesses formulated in an ideal situation without noise may not be able to effectively detect quantum correlations in the presence of noise.  

Here we introduce a noise-adaptive method to detect quantum correlations in terms of the generalized quasiprobability function. We analyze the effects of measurement imperfection and thermal effect by environment on quantum correlations in phase space. It is shown that their noise effects can be well encapsulated into the change of the order parameter of the generalized quasiprobability function. In this formalism, we formulate a noise-adaptive witness in the form of a Bell-type inequality by the generalized quasiprobability function. A violation of the proposed inequality is a direct indication of the existence of entanglement in the state of the measured system. Remarkably, it allows us to observe quantum correlations in a CV system even under a significant amount of noise. Our work provides a useful tool to test quantum correlations in near-term noisy quantum devices with CV systems.

\section{Generalized quasiprobability function}

We start with briefly reviewing the generalized representation of the quasiprobability distribution. As Cahill and Glauber  introduced in Refs.~\cite{Cahill69a,Cahill69b,Moya93}, the generalized parity operator defined by
\begin{equation}
\label{eq:gpo}
\hat{\Pi}(\alpha; s) = \frac{1}{1-s} \sum^{\infty}_{n=0}\bigg(\frac{s+1}{s-1}\bigg)^n \ket{\alpha,n}\bra{\alpha,n}
\end{equation}
forms a complete set of operators so that every bounded operator (i.e., the Hilbert-Schmidt norm of which is finite) can be represented by its expansion with appropriate weight functions. Here, $\ket{\alpha,n}$ is the displaced number state produced by applying the Glauber displacement operator $\hat D(\alpha)$ to the number basis, $\ket{\alpha,n}\equiv\hat D(\alpha)|n\rangle$ with a complex parameter $\alpha$. As every density matrix of a quantum state $\hat{\rho}$ is bounded, we can represent arbitrary quantum state in terms of the weight function given as the expectation value of $\hat{\Pi}(\alpha; s)$,
\begin{eqnarray}
\label{eq:sQP} 
W(\alpha; s)=\frac{2}{\pi}\mathrm{Tr}[\hat{\rho}\hat{\Pi}(\alpha; s)],
\end{eqnarray}
which is the so called $s$-parametrized quasiprobability function \cite{Cahill69a,Cahi$P$-funll69b}. This is the unified form of quasiprobability functions with different orders parameter $s$: (i) If $s$ tends to one from the left, it becomes the Glauber-Sudarshan $P$-function \cite{Sudarshan,Glauber63},
\begin{equation}
P(\alpha)=\lim_{s\rightarrow 1-}\frac{2}{\pi}\mathrm{Tr}[\hat{\rho}\hat{\Pi}(\alpha; s)].
\end{equation}
Note that the expectation value of $\hat{\Pi}(\alpha; s)$ for $\ket{\alpha,n}$ gets infinity as $s\rightarrow1$, representing the singularity of the $P$-fuparametriznction \cite{Cahill69a,Cahill69b}. (ii) If we set $s=0$, $\hat{\Pi}(\alpha; 0) = \sum^{\infty}_{n=0}(-1)^n \ket{\alpha,n}\bra{\alpha,n}=\hat D(\alpha)(-1)^{\hat{n}}\hat D(\alpha)$, and the Wigner function \cite{Wigner}
\begin{equation}
W(\alpha)=\frac{2}{\pi}\mathrm{Tr}[\hat{\rho}\hat{\Pi}(\alpha; 0)]=\frac{2}{\pi} \sum^{\infty}_{n=0}(-1)^n\bra{\alpha,n}\rho\ket{\alpha,n}
\end{equation}
is obtained. (iii) If $s=-1$, $\hat{\Pi}(\alpha; -1) = \ket{\alpha}\bra{\alpha}$ so that the Eq.~(\ref{eq:sQP}) becomes the Husimi $Q$-function \cite{Husimi} as
\begin{equation}
Q(\alpha)=\frac{1}{\pi}\mathrm{Tr}[\hat{\rho}\hat{\Pi}(\alpha; -1)]=\frac{1}{\pi} \bra{\alpha}\rho\ket{\alpha}.
\end{equation}
\begin{figure}
\centering
\includegraphics[width=0.7\linewidth]{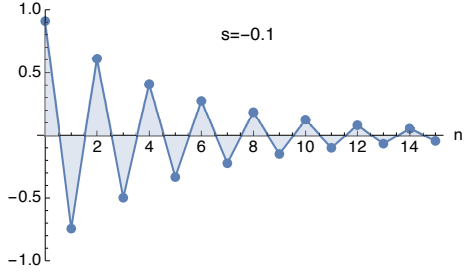}
\caption{Change of the expectation value of the generalized parity operator $\hat{\Pi}(\alpha; s)$ with $n$ for a displaced number state $\ket{\alpha,n}$ when $s=-0.1$.}\label{fig:eigen}
\end{figure}

In general, a quasiprobability distribution $W(\alpha; s')$ can be regarded as a smoothed quasiprobability distribution of $W(\beta; s)$ with an order parameter $s>s'$. This can be represented by the convolution of $W(\beta; s)$ and a Gaussian weight \cite{Cahill69a,Cahill69b}
\begin{equation}
\label{eq:convG}
 W(\alpha; s') = \frac{2}{\pi (s-s')}\int d^2\beta
~W(\beta; s)\exp\left(-{\frac{2|\alpha-\beta|^2}{s-s'}}\right).
\end{equation}
As the effects of noise in the phase space representation can be modeled by Gaussian smoothing \cite{pndist,Leonhardt93,Banaszek96,Banaszek99a,Banaszek02,Lvovsky2001,Thomas00,HWLee15}, the decreasing $s'$ is often considered as a loss of non-classicality. For example, the negativity of $W(\beta; s)$ is reduced as $s'$ decreases, and the expectation value becomes non-negative everywhere in phase space when it becomes the Husimi $Q$-function $s'=-1$. This can be understood as the smoothing of $W(\beta; s)$ over the area satisfying the Heisenberg minimum uncertainty, associated with the ideal simultaneous measurements of position and momentum. Hence, the Husimi $Q$-function can be often interpreted as a proper probability distribution. 

The expectation value of the generalized parity operator $\hat{\Pi}(\alpha; s)$ for a displaced number state $\ket{\alpha,n}$ is bounded when $s\leq0$, while it diverges when $s>0$. When $s=0$, the expectation value is given as $(-1)^n$. When $s<0$, the absolute value of the expectation value is always less than 1 and becomes smaller as increasing $n$. For example, the change of the expectation value for $\ket{\alpha,n}$ with $s=-0.1$ is plotted in Fig.~\ref{fig:eigen}. We intend to use $\hat{\Pi}(\alpha; s)$ as an observable to test quantum correlations so that we shall focus on the non-positive $s$ region in the later part of this paper. Then, the $s$-parametrizd quasiprobability function $W(\alpha; s)$ covers from the Wigner function with $s=0$ to the Husimi $Q$-function with $s=-1$.

\section{Effect of detection noise}

For the number resolving detection of bosonic particles (e.g., photons), the noise effect is mainly due to the loss of particles by inefficient detectors. Such imperfection can be modeled by the action of a beam-splitter with a transmittance $\eta$. We do not consider here the dark counts since their effects are relatively minor than losses especially when the detection efficiency is low. The probability of the detection of each single particle is then $\eta$, and the overall effect on the number distribution can be described by the Bernoulli sampling. If the probability that the field contains $n$ particles is $p(n)$, the probability for detecting $m$ particles is given by \cite{pndist}
\begin{equation}
\label{eq:Berdis}
p_{\eta}(m)=\sum^{\infty}_{n=m}\binom{n}{m}(1-\eta)^{n-m}\eta^m p(n),
\end{equation}
where $\binom{n}{m}=n!/[m!(n-m)!]$ is the binomial coefficient.

The quasiprobability distribution reconstructed by inefficient detectors with $\eta$ is then written by 
\begin{equation}
\label{eq:qdcoarse}
W_{\eta}(\alpha;s)=\frac{2}{\pi(1-s)}\sum^{\infty}_{m=0}\Big(\frac{s+1}{s-1}\Big)^mp_{\eta,\alpha}(m),
\end{equation}
where $p_{\eta,\alpha}(m)=\sum^{\infty}_{n=m}\binom{n}{m}(1-\eta)^{n-m}\eta^m p_{\alpha}(n)$ and $p_{\alpha}(n)\equiv\bra{\alpha,n}\rho\ket{\alpha,n}\equiv\bra{n}\hat{D}(\alpha)\rho\hat{D}(\alpha)\ket{n}$. This can be then recast into (see appendix)
\begin{eqnarray}
 \label{eq:Wexp}
 \nonumber
  W_{\eta}(\alpha; s)&=&\frac{2}{\pi(1-s)}\sum^{\infty}_{n=0}\Big(1-\eta+\eta\frac{s+1}{s-1}\Big)^n P(\alpha,n)\\
  &=&\frac{1}{\eta}W\bigg(\alpha; -\frac{1-s-\eta}{\eta} \bigg) \equiv  \frac{1}{\eta}W(\alpha; s').
\end{eqnarray}
Consequently, the generalized quasiprobability function measured with detection efficiency $\eta$ can be identified  with the quasiprobability function with a rescaled parameter 
\begin{eqnarray}
\label{eq:efficiencys} 
1-s'=\frac{1-s}{\eta}.
\end{eqnarray}
Note that the relation in Eq.~(\ref{eq:efficiencys}) is generally valid for any reconstruction method of the quasiprobability distribution. For example, the result is consistent with the analysis of the noise effects when reconstructing the quasiprobability distribution by homodyne detection \cite{pndist,Leonhardt93, Lvovsky2001,Thomas00,HWLee15}.

\section{Effect of noise from environment}

Let us first introduce the convolution law of the quasiprobability distribution function \cite{MSKim95}. Consider a beam splitter with the transmissivity $t$ and reflectivity $r$ satisfying $r^2+t^2=1$. The $Q$-function ($s=1$) of one output mode (denoted by mode $d$) is the simple convolution of the two input modes (denoted by $a$ and $b$ modes) \cite{Cahill69a,Cahill69b} so that the characteristic functions of the input and output modes are in the convolution relation as $\chi_{d}(\alpha; s=1)=\chi_{a}(r\alpha; s=1)\chi_{b}(t\alpha; s=1)$. As the generalized characteristic function of a quantum state $\hat{\rho}$ with an order parameter $s$ can be written by
$\chi(\alpha;s)=\mathrm{Tr}[\hat{\rho}\exp(\alpha\hat{a}^{\dag}-\alpha^{*}\hat{a})]\exp(s|\alpha|^2/2)=\chi(\alpha;1)\exp[(s-1)|\alpha|^2/2]$, we can obtain the convolution law for the generalized characteristic function as
\begin{eqnarray}
\label{eq:sconv} \nonumber
\chi_{d}(\alpha;s)&=&\chi_{d}(\alpha;1)\exp\bigg(\frac{s-1}{2}|\alpha|^2\bigg)\\
\nonumber
&=&\chi_{a}(r\alpha;1)\exp\bigg(\frac{s-1}{2}|r\alpha|^2\bigg)\\
&&~~~~\times
\chi_{b}(t\alpha;1)\exp\bigg(\frac{s-1}{2}|t\alpha|^2\bigg)\nonumber\\
&=&\chi_{a}(r\alpha;s)\chi_{b}(t\alpha;s).
\end{eqnarray}
As a result, we can arrive at the convolution law for the generalized quasiprobability distribution function,
\begin{eqnarray}
\label{eq:sconvQuasi} W_{d}(\alpha; s)&=&\frac{1}{t^2}\int d^2\beta
W_{a}(\beta; s)W_{b}\Big(\frac{\alpha-r \beta}{t};s\Big).
\end{eqnarray}

Let us now analyze the effect of noise due to the environment. Since a dissipation induced by interaction with environment tends to smooth the quasiprobability distribution, the state evolution under environmental noise can be effectively described by a dynamical change of the order parameter $s$. This approach is also in agreement with the description of the measurement imperfection given in the previous section, since an attenuated dynamics can be also understood within the framework of noisy measurements \cite{Buzek95}. As an exemplary model, we focus here on a noise model induced by thermal environment. Suppose that a quantum system encounters and interacts with thermal environment regarded as a reservoir. The effect of the reservoir can be modeled by mixture of the mode for the system and the mode of thermal fields by a beam splitter. The evolution of the quasiprobability distribution can be described by the solution of the Fokker-Planck equation \cite{MSKim95},
\begin{equation}
\begin{aligned}
\label{eq:FPeq} 
&\frac{\partial W(\alpha; s; \tau)}{\partial \tau}=\\
&~~\frac{\kappa}{2}\bigg[\frac{\partial}{\partial \alpha}\alpha+\frac{\partial}{\partial \alpha^*}\alpha^*+2\bigg(\frac{1}{2}+\bar{n}\bigg)\frac{\partial^2}{\partial \alpha \partial \alpha^*} \bigg]W(\alpha; s; \tau).
\end{aligned}
\end{equation}
We then obtain the time evolution at time $\tau$ by means of the convolution of the original field and thermal environment as
\begin{equation}
\label{eq:solFPeq}
W(\alpha;s;\tau)=\frac{1}{t^2(\tau)}\int d^2 \beta W^{\mathrm{th}}(\beta;s)W\bigg(\frac{\alpha-r(\tau)\beta}{t(\tau)};s;0\bigg),
\end{equation}
where $r(\tau)=\sqrt{1-e^{-\gamma\tau}}$ and $t(\tau)=\sqrt{e^{-\gamma\tau}}$ are given in terms of the energy decay rate $\gamma$. Here,
\begin{equation}
W^{\mathrm{th}}(\beta;s)=\frac{2}{\pi(1+2\bar{n}-s)}\exp{\biggl(-\frac{2|\beta|^2}{1+2\bar{n}-s}\biggr)}
\end{equation}
is the generalized quasiprobability function of the thermal state with mean photon number $\bar{n}$. By rescaling $\beta$ and $\alpha$ with respect to $\beta'=r(\tau)\beta/t(\tau)$ and $\alpha'=\alpha/t(\tau)$, Eq.~(\ref{eq:solFPeq}) can be recast into
\begin{equation}
\begin{aligned}
\nonumber
\frac{2}{\pi(1+2\bar{n}-s)r(\tau)^2}&\int d^2 \beta' W(\beta';s)\\
&\times\exp{\bigg(-\frac{2 t(\tau)^2 |\alpha'-\beta'|^2}{(1+2\bar{n}-s)r(\tau)^2}\bigg)}.
\end{aligned}
\end{equation}
From Eq.~(\ref{eq:convG}), the effect of thermal environment can then be identified with a temporal change of the quasiprobability function as
\begin{equation}
\label{eq:temporal}
W(\alpha;s;\tau)=\frac{1}{t^2(\tau)}W\bigg(\frac{\alpha}{t(\tau)};s'(\tau);0\bigg),
\end{equation}
with an order parameter given by
\begin{equation}
\label{eq:thermals}
s'(\tau)=\frac{s-r^2(\tau)(1+2\bar{n})}{t^2(\tau)}.
\end{equation}
Therefore, we have found that the evolution of quasiprobability distribution under the influence of thermal environment noise can be effectively described by a dynamical change of the order parameter $s$.

\section{Bell inequalities with quasiprobability functions}

In the following we formulate an entanglement witness in the form of a Bell-type inequality \cite{Bell64} in terms of the generalized quasiprobability function to test quantum correlations under noise. Suppose that a two-mode system is prepared to test. As the expectation value of the generalized parity operator $\hat{\Pi}(\alpha; s)$ is bounded with a non-positive $s$, we can use $\hat{\Pi}(\alpha; s)$ to define an observable for testing quantum correlations. Let us define the effective observable operator as
\begin{equation}
\label{eq:ObOp}
\hat{O}(\alpha;s)=X(s)\hat{\Pi}(\alpha;s)+Y(s)\openone,
\end{equation}
where $X(s)>0$ and $Y(s)$ are arbitrary functions of the order parameter $s$ in the region $-1\leq s\leq0$ and $\openone$ is the identity operator. Under the circumstance, the eigenvalue spectrum of the observable operator (\ref{eq:ObOp}) is given as
\begin{equation}
\label{eq:ObEigen}
e_n(s)=\frac{X(s)}{1-s}\bigg(\frac{s+1}{s-1}\bigg)^n+Y(s).
\end{equation}
It is straightforward to see that the value of $-(s+1)^n/(s-1)^{n+1}$ takes the maximum when $n=0$ and the minimum when $n=1$ in the region $-1\leq s\leq0$ (for example, see Fig.~\ref{fig:eigen}). The maximum and minimum eigenvalues of the operator are thus obtained as $e_0(s)=X(s)/(1-s)+Y(s)$ and $e_1(s)=-(s+1)X(s)/(1-s)^2+Y(s)$, respectively. 
If we assume that the observable operator (\ref{eq:ObOp}) is bounded by 
\begin{equation}
|\langle \hat{O}(\alpha;s) \rangle |\leq1,
\end{equation}
the conditions we can take are $X(s)/(1-s)+Y(s)=1$ and $-(s+1)X(s)/(1-s)^2+Y(s)=-1$. Hence, we arrive at a solution $X(s)=(1-s)^2$ and $Y(s)=s$. As a result, the operator has the form
\begin{equation}
\label{eq:ObOpform}
\hat{O}(\alpha;s)=(1-s)^2\hat{\Pi}(\alpha;s)+s\openone,
\end{equation}
with eigenvalues
\begin{equation}
e_n(s)=(1-s)\bigg(\frac{s+1}{s-1}\bigg)^n+s.
\end{equation}
Note that when $s=0$ it becomes $\hat{O}(\alpha;0)=\hat{\Pi}(\alpha; 0)=\sum^{\infty}_{n=0}(-1)^n\ket{\alpha,n}\bra{\alpha,n}$, and when $s=-1$, $\hat{O}(\alpha;-1)=2\ket{\alpha}\bra{\alpha}-\openone$. For the assignment $s=0$ and $s=1$, we can thus recover the Wigner and Husimi Q functions, respectively.

Now we formulate a Bell-type inequality in terms of the generalized quasiprobability function. Suppose that the two local parties choose observables, $\hat{A}_a$ and $\hat{B}_b$, where $a,b\in\{1,2\}$. The measurement operators of the local observables are defined as
\begin{equation}
 \label{eq:LOOp}
  \hat{A}_a =\hat{O}(\alpha_a; s),~~~~\hat{B}_a =\hat{O}(\beta_b; s),
\end{equation}
where $-1\leq s\leq0$. Let us then construct a Bell operator in a similar way with the CHSH-type \cite{CHSH69} as
\begin{equation}
 \label{eq:BOp}
  {\cal \hat{B}}= \hat{A}_1\otimes\hat{B}_1+\hat{A}_1\otimes\hat{B}_2+\hat{A}_2\otimes\hat{B}_1-\hat{A}_2\otimes\hat{B}_2.
\end{equation}
Note that the expectation value of each term of (\ref{eq:BOp}) for any separable states $\rho=\sum_i P_i \rho^a_i \otimes \rho^b_i$ with $\sum_i P_i=1$ is written by $\langle \hat{A}_a\otimes \hat{B}_b\rangle=\mathrm{Tr}[\hat{\rho}\hat{A}_a\otimes \hat{B}_b]=\sum_i P_i \mathrm{Tr}[\hat{\rho^a_i}\hat{A}_a]\mathrm{Tr}[\hat{\rho^b_i}\hat{B}_b]=\sum_i P_i \langle \hat{A}_a \rangle_i\langle \hat{B}_b \rangle_i$ so that the overall expectation value of the Bell operator is $\langle \hat{{\cal B}}\rangle=\sum_i P_i (\langle \hat{A}_1 \rangle_i\langle \hat{B}_1\rangle_i+\langle \hat{A}_1 \rangle_i\langle \hat{B}_2\rangle_i+\langle \hat{A}_2 \rangle_i\langle \hat{B}_1\rangle_i-\langle \hat{A}_2 \rangle_i\langle \hat{B}_2\rangle_i)$.
Since, in our case, all local observables are bounded by $|\langle \hat{A}_a \rangle_i|<1$ and $|\langle \hat{B}_b\rangle_i| \leq 1$ for any non-positive $s$, the expectation value of the Bell operator is bounded as $|\langle \hat{{\cal B}}\rangle| \equiv |{\cal B}| \leq 2$ by separable states.  

Finally, we can write a Bell function formulated by the generalized quasiprobability function as
\begin{widetext}
\begin{equation}
\begin{aligned}
\label{eq:BFinQP}
  |{\cal B}(s)|&=\big|\langle\hat{O}(\alpha_1; s)\otimes\hat{O}(\beta_1; s)\rangle+\langle\hat{O}(\alpha_1; s)\otimes\hat{O}(\beta_2; s)\rangle+\langle\hat{O}(\alpha_2; s)\otimes\hat{O}(\beta_1; s)\rangle-\langle\hat{O}(\alpha_2; s)\otimes\hat{O}(\beta_2; s)\rangle\big|\\
  &=\biggl|
  \frac{\pi^2(1-s)^4}{4}[W(\alpha_1,\beta_1;s)+W(\alpha_1,\beta_2;s)+W(\alpha_2,\beta_1;s)-W(\alpha_2,\beta_2;s)]\\
  &\hspace{75mm}+\pi s(1-s)^2[W(\alpha_1;s)+W(\beta_1;s)]+2s^2 \biggr|
  ~\leq 2,
\end{aligned}
\end{equation}
\end{widetext}
for $-1\leq s\leq0$, where $W(\alpha,\beta; s) =(4/\pi^2)\langle \hat{\Pi}(\alpha; s \otimes\hat{\Pi}(\beta; s)\rangle$ is the two-mode quasiprobability function and $W(\alpha;s)$ and $W(\beta;s)$ are the marginal distribution functions. Therefore, a violation of the inequality (\ref{eq:BFinQP}) guarantees that the state is entangled. Note that this is not a test of quantum non-locality, which has a different criterion in terms of a local realistic theory \cite{Bell64,CHSH69,Swlee2010}, but it is a method for witnessing entanglement of a CV system in phase space.

\section{Noise-adaptive test of quantum correlations}

Let us now apply the formulated witness to test quantum correlations of a CV system under noise. 
We shall consider a two-mode squeezed vacuum state (TMSV) as a representative example,
\begin{eqnarray}
\label{eq:TMSSs}
\ket{\mathrm{TMSV}}=\sum_{n=0}^{\infty}\frac{\tanh^{n}{\xi}}{\mathrm{cosh}~\xi}\ket{n,n},
\end{eqnarray}
with a squeezing parameter $\xi>0$. TMSW can be generated, e.g., by non-degenerate optical parametric amplifiers \cite{Reid88}, and has often been regarded as the normalized EPR states, i.e., the maximally entangled CV state associated with position and momentum \cite{Banaszek99}. For a non-positive parameter $s$, its generalized quasiprobability distribution function is given by
\begin{equation}
\begin{aligned}
 \label{eq:QpfTMSS}
W(\alpha,\beta;s)&=\frac{4}{\pi^2R(s)}\exp\biggl(-\frac{2}{R(s)}\{S(s)(|\alpha|^2+|\beta|^2)\\
  &+\sinh{2\xi}(\alpha\beta+\alpha^*\beta^*)\}\biggr),
  \end{aligned}
\end{equation}
where $R(s)=s^2-2s\cosh{2\xi}+1$ and $S(s)=\cosh{2\xi}-s$, and its marginal single-mode distribution is $W(\alpha;s)=(2/\pi S(s))\exp[-2|\alpha|^2/S(s)]$. 

\subsection{Quantum correlations under detection noise}

\begin{figure}
\includegraphics[width=1.0\linewidth]{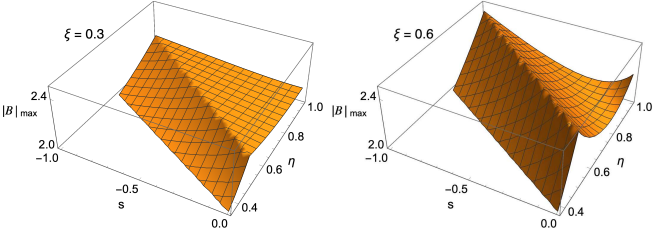}
\caption{The maximum expectation values of $|{\cal B}(s')|$ with TMSV are plotted in the range of $s$ and $\eta$ for different squeezing $\xi=0.3,~0.6$. Quantum correlations are detected by the violations of the inequality $|{\cal B}(s')|\leq 2$ with detection efficiency as low as about $\eta\sim0.36$.}\label{fig:BV1}
\end{figure}

We first consider a test of quantum correlations under detection noise. Assuming that the detection efficiencies in two modes are same as $\eta$, the reconstructed $s$-parametrizd quasiprobability functions for two modes and single mode are given respectively by
\begin{equation}
\begin{aligned}
W_{\eta}(\alpha,\beta; s)&=\frac{1}{\eta^2}W(\alpha,\beta; s')\\
W_{\eta}(\alpha; s)&=\frac{1}{\eta}W(\alpha; s').
\end{aligned}
\end{equation}
Since $\eta$ is assumed to be a known parameter here, we can represent the reconstructed distributions in terms of the quasiprobability function with a rescaled order parameter $s'$ given in Eq.~(\ref{eq:efficiencys}),
\begin{equation}
\begin{aligned}
W(\alpha,\beta; s')&=\eta^2 W_{\eta}(\alpha,\beta; s)\\
W(\alpha; s')&=\eta W_{\eta}(\alpha; s).
\end{aligned}
\end{equation}
Therefore, the Bell function with the reconstructed quasiprobability distributions and the rescaled order parameter $s'=s/\eta+(1-1/\eta)$ is given as
\begin{equation}
\begin{aligned}
\label{eq:effBFinQP}
 & |{\cal B}(s')|=\biggl|
  \frac{\pi^2(1-s')^4}{4}[W(\alpha_1,\beta_1;s')+W(\alpha_1,\beta_2;s')\\
  &~~~+W(\alpha_2,\beta_1;s')-W(\alpha_2,\beta_2;s')]+\pi s'(1-s')^2\\
  &~~~\times[W(\alpha_1;s')+W(\beta_1;s')]+2s'^2 \biggr|
  \\
  &=\biggl|
  \frac{\pi^2(1-s')^4\eta^2}{4}[W_{\eta}(\alpha_1,\beta_1;s)+W_{\eta}(\alpha_1,\beta_2;s)\\
  &~~~+W_{\eta}(\alpha_2,\beta_1;s)-W_{\eta}(\alpha_2,\beta_2;s)]+\pi s'(1-s')^2\eta\\
  &~~~\times[W_{\eta}(\alpha_1;s)+W_{\eta}(\beta_1;s)]+2s'^2 \biggr|~\leq 2,
\end{aligned}
\end{equation}
for $-1\leq s'\leq0$. We keep ${\cal B}(-1)$ for $s'<=-1$. Notably, here we choose the Bell inequality $|{\cal B}(s')|\leq2$ as a noise-adaptive witness by rescaling the order parameter as $s'=s/\eta+(1-1/\eta)$. This is in contrast to the approach in Ref.~\cite{SWLee09}, where the Bell inequality $|{\cal B}(s)|\leq2$ is used to test quantum correlations without changing the given order parameter $s$ even in the presence of noise. The inequality in Eq.(\ref{eq:effBFinQP}) can be considered as a generalized form of the noise-adaptive entanglement witness with the Wigner function proposed in Ref.~\cite{Swlee2010}.

In Fig.~\ref{fig:BV1}, we plot the violations of the Bell-type inequality in Eq.~(\ref{eq:effBFinQP}) for TMSV by changing $\eta$ and $s$ with different squeezing rate $\xi$. Remarkably, quantum correlations are observed even when the detection efficiency is as low as $\eta\sim0.36$ for $\xi=0.3$ and $\eta\sim0.37$ for $\xi=0.6$ when $s=0$. This is a significant improvement over the test with the other known entanglement witnesses, e.g., proposed in Refs.~\cite{Banaszek02,SWLee09,Swlee2010}, under the effect of noise.

We observe that the amount of violation shows different tendencies depending on $s$, $\eta$ and the squeezing parameter $\xi$. Peaks are observed at $s'=s/\eta+(1-1/\eta)=-1.0$. It shows that the dominant part of the violation in this region comes from the vacuum-photon entanglement, because the measurement operator in Eq.(\ref{eq:ObOpform}) becomes the photon on-off detection when $s'=-1.0$. Upon increasing $\xi$, a narrower peak of violation appears at the region $s=0$ and $\eta=1$, which is the detection of the entanglement between multiple photons of two modes. The observable operator in Eq.(\ref{eq:ObOpform}) becomes the parity operator when $s'=0$. Note that the parity measurement can detect the correlation between higher number of photons than the on-off measurement but is more fragile under detection noise.

\subsection{Dynamical quantum correlations under thermal environment}

We then test a dynamic behavior of quantum correlations under the effect of thermal environmental noise. We assume that the thermal noise in two modes is independent and has same energy decay rate $\gamma$ and average thermal photon number $\bar{n}$. Using the rescaled quasiprobability function in Eq.~(\ref{eq:temporal}) and the order parameter in Eq.~(\ref{eq:thermals}), the evolution of the quasiprobability distribution of TMSV can be represented in terms of
\begin{widetext}
\begin{equation}
\begin{aligned}
\label{eq:temporalTMSV}
W(\alpha,\beta;s;\tau)&=\frac{1}{t^4(\tau)}W\bigg(\frac{\alpha}{t(\tau)},\frac{\beta}{t(\tau)};s'(\tau);0\bigg)\\
&=\frac{4}{\pi^2t^4(\tau)R'(s,\tau)}\exp\biggl(-\frac{2}{R'(s,\tau)}\bigg\{S'(s,\tau)\frac{|\alpha|^2+|\beta|^2}{t^2(\tau)}+\sinh{2\xi}\frac{\alpha\beta+\alpha^*\beta^*}{t^2(\tau)}\bigg\}\biggr),
\end{aligned}
\end{equation}
and the marginal distribution
\begin{eqnarray}
\label{eq:temporalTMSVma}
W(\alpha;s;\tau)&=&\frac{1}{t^2(\tau)}W\bigg(\frac{\alpha}{t(\tau)};s'(\tau);0\bigg)=\frac{1}{\pi t^2(\tau) S'(s,\tau)}\exp\bigg(-\frac{2|\alpha|^2}{t^2(\tau)S'(s,\tau)}\bigg),
\end{eqnarray}
where $R'(s,\tau)=s'(\tau)^2-2s'(\tau)\cosh{2\xi}+1$ and $S'(s,\tau)=\cosh{2\xi}-s'(\tau)$ with the parameters $s'(\tau)=(s-r^2(\tau)(1+2\bar{n}))/t^2(\tau)$ and $r(\tau)=\sqrt{1-e^{-\gamma\tau}}$ and $t(\tau)=\sqrt{e^{-\gamma\tau}}$. Therefore, we can set a Bell-type inequality by rescaling with respect to the dynamically changing parameters $\alpha'=\alpha/t(\tau)$, $\beta'=\beta/t(\tau)$, and $s'(\tau)$ as
\begin{equation}
\begin{aligned}
\label{eq:effBFinQPD}
|{\cal B}(s'(\tau))|=&\biggl| \frac{\pi^2(1-s'(\tau))^4}{4}[W(\alpha'_1,\beta'_1;s'(\tau))+W(\alpha'_1,\beta'_2;s'(\tau))+W(\alpha'_2,\beta'_1;s'(\tau))-W(\alpha'_2,\beta'_2;s'(\tau))]\\
  &+\pi s'(\tau)(1-s'(\tau))^2 [W(\alpha'_1;s'(\tau))+W(\beta'_1;s'(\tau))]+2s'(\tau)^2 \biggr|\\
  =&\biggl|
  \frac{\pi^2(1-s'(\tau))^4 t(\tau)^4}{4}[W(\alpha_1,\beta_1;s;\tau)+W(\alpha_1,\beta_2;s;\tau)+W(\alpha_2,\beta_1;s;\tau)-W(\alpha_2,\beta_2;s;\tau)]\\
  &+\pi s'(\tau)(1-s'(\tau))^2 t(\tau)^2[W(\alpha_1;s;\tau)+W(\beta_1;s;\tau)]+2s'(\tau)^2 \biggr|~\leq 2.
\end{aligned}
\end{equation}
\end{widetext}

\begin{figure}
\includegraphics[width=1.0\linewidth]{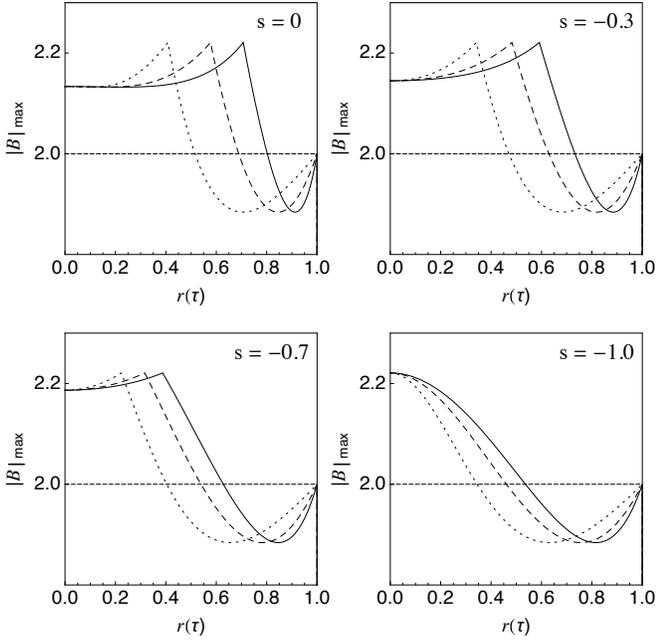}
\caption{The time evolutions of the maximum expectation values of $|{\cal B}(s')|$ with TMSV under the effect of thermal environment is plotted against the dimensionless time $r(\tau)$, which is $0$ when $\tau=0$ and $1$ when $\tau=\infty$ for different order parameter $s$ with squeezing $\xi=0.3$. The average photon number of the thermal environment is $\bar{n}=0$ (solid line), $\bar{n}=0.5$ (dashed line), and $\bar{n}=2$ (dotted line).}\label{fig:BV2}
\end{figure}

In Fig.~\ref{fig:BV2}, we plot the dynamics of quantum correlations of TMSV detected by the witness in Eq.~(\ref{eq:effBFinQPD}) under thermal environmental noise. Remarkably, it is possible to observe quantum correlations of TMSV ($\xi=0.3$) up to the dimensionless time $r(\tau)\sim 0.8,~0.7,~0.5$ under thermal environment with the average photon number $\bar{n}=0,~0.5,~2$, respectively. Note that these are much longer than the time for which one can detect quantum correlations by previous schemes. For example, by the scheme in Ref.~\cite{Jeong2000}, quantum correlations can be observed up to the time $r(\tau)\sim 0.35,~1.3,~0.6$ under the same thermal environment noise with $\bar{n}=0,~0.5,~2$, respectively.

\section{Remarks} 

As proposed in the previous investigation, the formalism can be generalized further for testing high-dimensional quantum correlations. The observable operator $\hat{\Pi}(\alpha; s)$ in Eq.~(\ref{eq:gpo}) for $s\leq 0$ can be associated with a noisy measurement process performed by a dichotomic (two-dimensional) binning into the outcomes $\pm1$ after number-resolving detection. Similarly, we can map the number $n$ into the discretized phases by $\omega = \exp({2\pi i/d})$ for a measurement with arbitrary $d$ outcomes. The eigenvalue of the observable can be assigned as a complex variable $\omega^{n}$. Therefore, the generalized quasiprobability function with $d$-dimensional outcomes can be defined as
\begin{equation}
\label{eq:gpd}
\begin{aligned}
W(\alpha; s_d)&=\frac{2}{\pi(1-s_d)} \sum^{\infty}_{n=0}\omega^n\bra{\alpha,n}\rho\ket{\alpha,n}\\
&=\frac{2}{\pi}\mathrm{Tr}[\hat{\rho}\hat{\Pi}(\alpha; s_d)],
\end{aligned}
\end{equation}
as the expectation value of the generalized parity operator 
\begin{equation}
\label{eq:gpod}
\hat{\Pi}(\alpha; s_d) = \frac{1}{1-s_d}\sum^{\infty}_{n=0}\bigg(\frac{s_d+1}{s_d-1}\bigg)^n \ket{\alpha,n}\bra{\alpha,n},
\end{equation}
with a complex order parameter $s_d=-i\cot(\pi/d)$. Note that Eq.~(\ref{eq:gpd}) becomes equivalent to Eq.~(\ref{eq:sQP}) for $d=2$. The operator in Eq.~(\ref{eq:gpod}) is associated with a measurement process performed by a number resolving detection and a subsequent binning into the complex value $\omega^{n}$. By using Eq.~(\ref{eq:gpd}), different type of Bell inequalities can be tested with arbitrary $d$-outcome measurements \cite{Collins02,WSon05,SWLee07,SWLee09r,Bae18}. For example, see the results in Refs.~\cite{WSon06,Swlee2011}. We can rewrite the $d$-dimensional quasiprobability function under noise as,
\begin{equation}
 \label{eq:Wexpd}
W_{\eta}(\alpha; s_d)=\frac{2}{\pi(1-s_d)}\sum^{\infty}_{n=0}(1-\eta+\eta\omega)^n P(\alpha, n)\equiv  \frac{W(\alpha; s'_d)}{\eta}.
\end{equation}
Note that the relation in Eq.~(\ref{eq:efficiencys}) is also valid here as 
\begin{eqnarray}
\label{eq:efficiencysd} 
1-s'_d=\frac{1-s_d}{\eta}.
\end{eqnarray}
Therefore, it would be possible to test high-dimensional quantum correlations under noise likewise the method proposed here. It may also be valuable to extend this method for testing multi-mode quantum correlations of CV systems in phase space \cite{vanLocck00,vanLocck01,SWLee13}.

Through series of studies in recent years, a hypothesis on the reason why quantum correlations disappear at macroscopic scale has been suggested based on coarse-graining measurements \cite{Kofler2007,Kofler2008}. A detection of quantum correlation with an extremely coarse-graining measurement was reported \cite{Jeong2009} but has been also understood within this hypothesis, because its local measurement requires the Kerr nonlinearity implicating the use of a precision measurement. A recent observation of the difficulty in detecting micro-macro entanglement by coarse-graining measurements \cite{Raeisi2011} has strengthen the validity of this hypothesis. However, our result clearly shows the possibility of direct observation of quantum correlations by coarse-graining measurements when the noise causing the coarse-graining can be identified. It implicates that quantum correlations may not entirely disappear in a macroscopic system but may be hidden to some extent regardless of the effect of noise. Therefore, our approach may provide an alternative way to explore the border between quantum and classical at macroscopic scale as well as to circumvent the difficulty in observing quantum correlations in complex systems.

The proposed witness is formulated as a CHSH-type inequality so that its capability to detect quantum correlations is inherited from the CHSH-type Bell inequality \cite{Banaszek99,SWLee09,CHSH69}. 
Therefore, it may be also valuable to analyze other criteria, e.g., the Peres-Horodecki criterion for CV systems \cite{Simon} which is more efficient to test weakly entangled states, based on our noise-adaptive approach.

In summary, we have proposed a method for testing quantum correlations by quasiprobability functions in the presence of noise. We have investigated the effects of measurement imperfection and thermal environmental noise on quantum correlations and shown that they can be encapsulated into the change of the order parameter of the generalized quasiprobability function. We have then formulated a noise-adaptive witness of quantum correlations in the form of a Bell-type inequality. Remarkably, it has been shown that the proposed witness allows us to detect quantum correlations in CV systems under a significant amount of noise. As the scheme is proposed based on current photonic detection technologies, an immediate experimental demonstration is expected. We believe that our method provides a useful tool to test quantum correlations in various protocols in near-term noisy quantum processors with CV systems.

\acknowledgments

SWL and JK were supported by KIAS Advanced Research Program (QP029902 and CG014604). W.S. supported by the Visiting Professorship Program at KIAS and Samsung Research Funding \& Incubation Center of Samsung Electronics under Project No. SRFC-IT1901-09.

\appendix*

\section{}

Here, we would like to show how the noise factor $\eta$ can be translated into the order parameter $s$. We can rewrite Eq.~(\ref{eq:qdcoarse}) with respect to the probability $p_{\alpha}(n)$ by Eq.~(\ref{eq:Berdis}) as
\begin{eqnarray}
 \nonumber
 &&\sum^{\infty}_{m=0}\left(\frac{s+1}{s-1}\right)^m
 \sum^{\infty}_{n=m}\binom{n}{m}(1-\eta)^{n-m}\eta^m p_{\alpha}(n)\\
 \nonumber
 &=&\sum^{\infty}_{n=0}(1-\eta)^n
 \sum^{\infty}_{m=0}\left(\frac{(s+1)\eta}{(s-1)(1-\eta)}\right)^m\binom{n}{m}p_{\alpha}(n)\\
 \nonumber
 &=&\sum^{\infty}_{n=0}(1-\eta)^n
 \left(1+\frac{(s+1)\eta}{(s-1)(1-\eta)}\right)^np_{\alpha}(n)\\
 \nonumber
 &=&\sum^{\infty}_{n=0}\left(1-\eta+\eta\frac{s+1}{s-1}\right)^nP_{\alpha}(n),
\end{eqnarray}
where we used $\sum^{\infty}_{n=m}\binom{n}{m}=
\sum^{\infty}_{n=0}\binom{n}{m}$ as $\binom{m-1}{m}=\cdots=\binom{0}{m}=0$ and the
relation $\sum^{\infty}_{m=0}x^{m}\binom{n}{m}=(1+x)^n$.

\end{document}